\documentstyle[aps,prl,twocolumn]{revtex}

\draft

\begin{document}


\twocolumn[
\hsize\textwidth\columnwidth\hsize\csname@twocolumnfalse\endcsname

\title{GROUND STATE ENERGY OF THE LOW DENSITY BOSE GAS}
\author{Elliott H. Lieb$^{1}$ and Jakob Yngvason$^2$}
\address{$1.$ Department of Physics, Jadwin Hall,
Princeton University, P.~O.~Box 708, Princeton, New Jersey 08544\\
$2.$ Institut f\"ur Theoretische Physik, Universit\"at Wien,
Boltzmanngasse 5, A 1090 Vienna, Austria\\}
\date{27 October, 1997, }
\maketitle

\begin{abstract}
Now that the properties of  low temperature Bose gases at low density,
$\rho$, can be examined experimentally it is appropriate to revisit
some of the formulas deduced by many authors 4-5 decades ago.  One of
these is that the leading term in the energy/particle is $2\pi \hbar^2
\rho a/m$, where $a$ is the scattering length.  Owing to the delicate
and peculiar nature of bosonic correlations, four decades of research
have failed to establish this plausible formula rigorously. The only
known lower bound for the energy was found by Dyson in 1957, but it was
14 times too small.  The correct bound is proved here.

\end{abstract}

\vfill
\twocolumn
\vskip.5pc ]
\narrowtext

\def\x{{\bf x}}
\def\X{{\bf X}}
\def\R{{\bf R}}

With the renewed experimental interest in low density, low temperature
Bose gases, some of the formulas posited four and five decades ago have
been dusted off and re-examined.  One of these is the leading term in
the ground state energy. In the limit of small particle density, $\rho$,
\begin{equation} e_0( \rho) \approx \mu \ 4\pi \rho  a, 
\label{rho}
\end{equation} 
where  $e_0(\rho)$ is the ground state energy (g.s.e.)
per particle in  the thermodynamic limit, $a$ is the scattering length
(assumed positive) of the two-body potential $v$ for bosons of mass $m$,
and $\mu \equiv \hbar^2/2m$.

Is (\ref{rho}) correct? In particular, is it true for the hard-sphere
gas? While there have been many attempts at a rigorous proof of
(\ref{rho}) in the past forty years, none has been found so far. Our
aim here is to supply that proof for finite range, positive potentials.
As remarked below, (\ref{rho}) cannot hold unrestrictedly;
more than $a>0$ is needed.

An upper bound for $e_0( \rho)$ agreeing with (\ref{rho}) is not easy
to derive, but it was achieved for hard spheres by a variational
calculation \cite {FD}, which can be extended to include general,
positive potentials of finite range.  What remained unknown was a good
lower bound. The only one available is Dyson's \cite {FD}, and that is
about {\it fourteen times smaller} than (\ref{rho}). In this paper we
shall provide a lower bound of the desired form, and thus prove
(\ref{rho}). We can also give explicit error bounds for small enough
values of the dimensionless parameter $Y\equiv 4\pi \rho a^3/3$:
\begin{equation}
e_0(\rho) \geq \mu\ 4\pi  \rho a (1- CY^{1/17})
\label{error}
\end{equation}
for some fixed $C$ (which is not evaluated explicitly because $C$ and
the exponent $1/17$ are only of academic interest).  The bound
(\ref{error}) holds for {\it  all non-negative, finite range,
spherical, two-body potentials.} A typical experimental value
\cite{TRAP} is $Y\approx 10^{-5}$. Dyson's upper bound is $\mu\ 4\pi
\rho a (1+2Y^{1/3})(1-Y^{1/3})^{-2}$.

We conjecture that (\ref{rho}) requires only a positive scattering
length {\it and} the absence of any many-body, negative energy bound
state.  If there are such bound states then (\ref {rho}) is certainly
wrong, but this obvious caveat does not seem to have been clearly
emphasized before.  There is a `nice' potential with positive scattering
length, no 2-body bound state, but with a 3-body bound state \cite
{BA}.

Our method also obviously applies to the positive temperature free
energy (because Neumann boundary conditions give an upper bound to the
solution to the heat (or Bloch) equation).

We also give some explicit bounds for {\it
finite} systems, which might be  useful for experiments with traps, but
we concentrate here on the thermodynamic limit for simplicity.
For traps with slowly varying confining potentials, $V_{\rm ext}$, our
method will prove that the leading term in the energy is
given by the well known local density approximation \cite{LDA}, which
minimizes the gaseous energy (\ref{rho}) plus the confining energy,
with respect to $\rho({\bf x})$, 
namely, 
$$
{\cal E}(\rho) \equiv \int (V_{\rm ext}({\bf x}) \rho({\bf x}) +\mu 4 \pi a 
\rho({\bf x})^2)d^3{\bf x}
$$
is minimized subject to $\int \rho = N=$
number of particles. 

The fact that Dyson's lower bound was not improved for four decades,
despite many attempts, attests to the fact that bosons are subtle
quantum mechanical objects which can have peculiar correlations unknown
to fermions. {}For example, there is the  non-thermodynamic $N^{7/5}$ law
for the charged Bose gas that was discovered by Dyson \cite{FD2},
confirmed only 20 years later \cite {CL}, and which defies any simple
physical interpretation.  

The first understanding of (\ref{rho}) goes back to Bogoliubov
\cite{BO}, who also introduced the notion of `pairing' in Helium (which
resurfaced in the BCS theory for fermions).  Later, there were several
derivations of (\ref{rho}) (and higher order) \cite{Lee-Huang-YangEtc},
\cite{EL3a}.  The method of the pseudopotential, which is an old idea
of Fermi's, was closest to the Bogoliubov analysis. The `exact'
pseudopotential was constructed in \cite{EL1a}, but it did not help to
make this appealing idea more rigorous. Most of the derivations were in
momentum space, the exception being \cite{EL3a}, which works directly
in physical space and which can handle both long and short
range potentials.  See \cite{EL2} for a review. All these methods rely
on special assumptions about the ground state (e.g., selecting special
terms in a perturbation expansion, which likely diverges) and
it is important to derive a fundamental result like (\ref{rho}) without
extra assumptions.

In all of this earlier work one key fact was not understood, or at
least not clearly stated in connection with the derivation of
(\ref{rho}).  It is that there are two different regimes, even at low
density, with very different physics, even though the simple formula
(\ref{rho}) seems to depend only on the scattering length.  Recall that
the (two-body) scattering length is defined, for a spherically
symmetric potential, $v$, by
\begin{equation}
-\mu u_0^{\prime\prime}(r) +\frac{1}{2} v(r)u_0(r) =0,
\label{scat}
\end{equation}
with $u_0(0)=0$, $u_0(r)>0$ (which is equivalent to the absence of
negative energy bound states, and which is true for nonnegative $v$).
As $r\to \infty$, $u(r)\approx r-a$. (Note the $v/2$ and not $v$ in
(\ref {scat}) because of the reduced mass.) Thus, $a$ depends on $m$ 
in a nontrivial way, and there are two extremes:

1. {\it Potential Energy Dominated Region:} The hard sphere ($v(r) =
\infty$ for $r < a$), is the extreme case here; the scattering length
is independent of $m$, and the energy is mostly (entirely) {\it
kinetic}. We see this from (\ref{rho}) because $-m~\partial e_0/
\partial m $ is the kinetic energy (Hellmann-Feynman theorem). In this
regime the potential is so dominant that it forces the energy to be
mostly kinetic. The g.s. wave function is highly correlated.

2. {\it Kinetic Energy Dominated Region:} The typical case is a very
`soft' potential. Then $a\approx (m/\hbar^2)\int_0^{\infty} v(r)r^2
dr$, which implies, from (\ref{rho}), that $e_0$ hardly depends on
$m$.  Thus, the energy is almost all {\it potential.}  The g.s. wave 
function is essentially the noninteracting one in this limit.

In other words, `scattering length' is not a property of $v$ alone, and
the low density gas, viewed from the perspective of the bosons, looks
quite different in the two regimes. Nevertheless, as (\ref{rho}) says,
the energy cannot distinguish the two cases. Whether Bose-Einstein
condensation itself can notice the difference remains to be seen. 
Condensation will not be touched upon here, except to note that so far
{\it the only case with 2-body interactions in which B-E condensation has
been rigorously established is hard core lattice bosons, but only at
half filling} \cite{KLS}.

Dyson  \cite {FD} effectively converted
region 1 into region 2. We shall make use of his important idea, 
which substitutes a very soft potential for the original one (even
a hard core) at the price of sacrificing the kinetic energy,

We assume that the $N$ particles are in a $L\times L\times L$ cubic box,
$\Omega$.  The particle density is then $\rho = NL^{-3}$.  It is well
known that the energy per particle in the thermodynamic
limit, $e_0(\rho)$, does not depend on the details of (reasonable)
$\Omega$, so we are free to use a cube and take $N\to \infty$ through
any sequence we please, as far as $e_0(\rho)$ is concerned.  
We set $N=kM$ with $k$ an integer and $M$ the
cube of an integer, because we shall want to divide up $\Omega$ into $M$
smaller cubes (called {\it cells}) of length $\ell=(k/\rho)^{1/3}$. We
will take $M\to \infty$ with $\ell$ and $k=\rho \ell^3$ fixed, but large.

The N-body Schr\"odinger operator is 
\begin{equation}
H= -\mu
\sum_{i=1}^{N} ~ \Delta_i ~ + ~ 
\sum_{1 \leq i < j \leq N} ~ v(\x_i - \x_j)~.
\label{ham}
\end{equation}

{}For boundary conditions we impose Neumann (zero derivative)
boundary conditions on $\Omega$.  It is well-known that Neumann boundary
conditions give the lowest possible g.s.e.  for $H$, and hence its use
is appropriate for a discussion of a {\it lower} bound for the g.s.e.  
Denote this Neumann g.s.e. by $E_0(N,L)$.

Now divide $\Omega$ into $M$  cells and impose Neumann conditions on
each cell, which, as stated before, lowers the energy further.  We also
neglect the interaction between particles in different cells; this,
too, can only lower the energy because $v\ge 0$.

A lower bound for $E_0 ( N,L)$ is obtained by distributing
the $N$ particles in the $M$ cells and then finding a lower bound for
the energy in these cells, which are now independent. We then add these
$M$ energies. {}Finally, we minimize the total energy over {\it all
choices} of the particle number in each cell (subject to the total
number being $N$).  Despite the independence of the cells, the latter
problem is not easy. In particular, something has to be invoked to 
make sure that we do not end up with some cells having too large a 
number of particles and some cells having  too few. 

With $L,N$ and $M=N/(\rho \ell^3)$ fixed, let $Mc_n$, for $n=0,1,2,
\ldots$ denote the number of cells containing exactly $n$ particles.
Then the particle number and cell number constraints are
\begin{equation}
\sum_{n \geq 0} ~ n c_n ~=~ k=\rho \ell^{3}, \qquad\quad 
\sum_{n \geq 0} ~ c_n ~ =1
\label{con}
\end{equation}
and our energy bound is 
\begin{equation}
 E_0 ( N, L ) ~ \geq~ M~\min~\sum_{n \geq 0} ~ c_n ~ E_0 ( n, \ell),
\label{lb}
\end{equation}
where the minimum is over all $c_n \ge 0$ satisfying (\ref{con}).

The minimization would be easy if we knew that $E_0
( n, \ell)$ (or a good lower bound for it) is convex in $n$, 
for then the optimum would be $c_n =
\delta_{n,k}$. This convexity is very plausible, but we cannot
prove it (except in the thermodynamic limit, where it amounts to
thermodynamic stability). What we do know instead is {\it
superadditivity:}
\begin{equation}
E_0( {n+n'}, \ell)\geq 
E_0( n,  \ell)+E_0( n',  \ell)
\label{superadd}
\end{equation}
for all $n$, $n'$, and this
turns out to be an adequate substitute for controlling 
the large $n$ terms in (\ref{lb}). 

Eq. (\ref{superadd}) is 
an immediate consequence of the positivity of the potential
and it is used as follows.
Suppose, provisionally, that  we have a lower bound of the form
\begin{equation}
E_0( n,  \ell) \ge K( \ell)~ n (n-1)  \qquad 
{\rm for}\ 0\le n \leq 4k,
\label{provis}
\end{equation}
with $K(\ell)$ independent of $n$ for $0\le n \leq 4k$.
In fact, we shall later prove that for small enough $\rho$ (and
hence small enough $k$) and suitable $\ell$, (\ref{provis}) holds with
\begin{equation}
K(\ell) \ge \mu~4\pi a~\ell^{-3} (1-C'~Y^{1/17}),
\label{kay}
\end{equation}
with $C'$ some constant. [However, the analysis we give now, leading to
(\ref{bound}), does not depend on this particular form of  $K(\ell)$.]

Split the sum in  (\ref{lb}) into two pieces:
$0\le n <4k$ and $4k \le n$. Let $t\equiv  \sum_{n <4k} nc_n \le k$, 
so that $k-t =  \sum_{n \ge 4k} nc_n $.
{}From  (\ref{provis})
and Cauchy's  inequality (and $\sum_{n <4k} c_n \le 1$)
\begin{equation}
\sum_{n <4k} c_n E_0( n,  \ell) \ge K(\ell)~ 
t(t-1).
\label{lowersum}
\end{equation}
On the other hand, if $n\ge 4k$ then, by (\ref{superadd}), 
$E_0( n,  \ell) \ge (n/8k )E_0( 4k, \ell)$, so
\begin{equation}
\sum_{n \ge 4k} c_n E_0( n,  \ell) \ge \frac{k-t}{2}K(\ell) 
 (4k-1).
\label{uppersum}
\end{equation}
Upon adding (\ref{lowersum}) and (\ref{uppersum}) the factor
$t(t-1)+(k-t)(2k-1/2)$ is obtained. Although the number $t$ is unknown, 
we note that
this factor  is monotone decreasing
in $t$ in the interval $0\le t\le k$ ( which is where $t$ lies, 
by (\ref{con})). Thus, we can  set $t=k$ and obtain the same bound
as if we had convexity, i.e.,
\begin{equation}
E_0(N,L) \ge N~K(\ell)~(\rho \ell^3-1).
\label{bound}
\end{equation}

In summary, if we can show (\ref{provis}) for a box of a {\it fixed}
size $\ell$, for all particle numbers up to $n=4\rho \ell^{3}$, then we
will have obtained our goal, (\ref{error}), in the thermodynamic limit
{\it provided} we can show that the $K$ in (\ref{provis}) satisfies
(\ref{kay}) with the constant $C'$ when $\ell$ is large compared to the
mean particle spacing, i.e., $\rho ~\ell^3 > C''~Y^{-1/17}$. Then the
$C$ in (\ref{error}) is equals $C'+C''$.  

We now focus on a single cell and denote the $n$ coordinates
$(\x_1,...,\x_n)$ collectively by $\X$.
The first step in proving (\ref{kay}) is to replace the 
total potential, $\sum_{i<j} v(\x_i - \x_j)$,
by a lesser 
quantity, ${\cal W}_v(\X)$, the {\it nearest neighbor potential} 
defined by
\begin{equation}
{\cal W}_v(\X)\equiv \frac{1}{2}\sum_{i=1}^n v(\x_i-\x_{j(i)}),
\label{nnpot}
\end{equation}
where $j(i)$ is the nearest neighbor to particle $i$ in the
configuration $\X$. I.e., particle $i$ `feels' only its nearest
neighbor.  Hence, we  replace $H$ by the smaller operator
\begin{equation}
\widetilde  H\equiv  {\cal T}+{\cal W}_v \leq H,
\end{equation}
where ${\cal T}=-\mu \sum \Delta_i$ is the kinetic energy in
(\ref{ham}).  Since $v\geq0$,  the g.s.e.  of
$\widetilde  H$ satisfies $\widetilde  E_0(n,\ell)\leq E_0 (n,\ell)$.

To get into the kinetic energy dominated region, we wish to replace $v$
in (\ref{nnpot}) by a gentler potential $U$. To this end we generalize
Lemma 1 of \cite{FD} and simplify its proof.  

{\it LEMMA 1: Let
$v(r)\geq0$ and $v(r)=0$ for $r>R_0$.  Let $U(r)\geq 0$ be any 
function
satisfying $ \int U(r)r^2dr \leq 1$ and  $U(r)=0$ for $r<R_0$. Let
${\cal B}\subset \R^3$ be star-shaped (convex suffices) with respect 
to $0$.
Then, for all functions $\phi$,}
\begin{equation}
\int_{\cal B} \mu|\nabla \phi (\x)|^2 + [\frac{1}{2}v(r)-\mu aU(r)]
|\phi (\x)|^2 d^3\x  \geq 0.
\label{dineq}
\end{equation}

{\it Proof:} Actually, (\ref{dineq}) holds with $\mu |\nabla \phi 
(\x)|^2$
replaced by the (smaller) {\it radial kinetic energy},
 $\mu |\partial \phi (\x)/ \partial r|^2$, and thus it  suffices to 
prove
the analog of (\ref{dineq}) for the integral along each radial
line, and to assume that $\phi(\x) = u(r)/r$ along this
line, with $u(0)=0$. Let us first prove
(\ref {dineq}) when $U$ is a delta-function at some radius $R\geq 
R_0$,
i.e., $U(r)=R^{-2}\delta (r-R)$. 
Then, it is enough to show, for all $u$, that 
\begin{equation}
 \int_0^R \mu |u'(r)-u(r)/r|^2 + \frac{1}{2}v(r)|u (r)|^2 dr
\geq \mu a|u(R)|^2 R^{-2}.
\label{oned}
\end{equation}
If the length of the radial line is 
less than $R$ then (\ref{oned}) is trivial. Otherwise,
normalize $u$ by $u(R)=R-a$, and ask for the minimum of the left side
of (\ref{oned}) under the condition that $u(0)=0$, $u(R)=R-a$. This is
a simple problem in the calculus of variations and leads to the
scattering length equation (\ref{scat}). If we substitute the solution
into (\ref{oned}), integrate by parts, and note that $u_0(r)=r-a$ for
$r>R_0$, we find that (\ref{oned}) is true if $a\leq R$,
which is true since $u_0\geq0$. {}Finally, by linearity and the fact 
that
$U(r)=\int r^{-2}\delta (r-s)U(s) s^2 ds$, the $\delta$-function case 
implies the general case. Q.E.D.

We select our $U$ by picking some $R \gg R_0$ and setting
\begin{equation}
U(r) ~= ~3(R^3-R_0^3)^{-1}  \qquad {\rm for}~ R_0<r<R
\label{yooo}
\end{equation}
and $U(r)=0$ otherwise. Later on we 
shall choose $R$, and we shall take 
\begin{equation}
R_0 \ll R \ll \rho^{-1/3} \ll \ell.
\end{equation}

By further decomposing a cube into Voronoi cells (which are always convex), 
Dyson 
\cite{FD} deduces from Lemma 1 that $\widetilde H$ is bounded below by 
a nearest neighbor potential, as in (\ref{nnpot}), i.e.,
\begin{equation}
H>\widetilde H> \mu a\  {\cal W}_U(\X),    
\label{lower}
\end{equation}
where ${\cal W}_U$ is as in (\ref {nnpot}), with $v$ replaced by $U$. 
For the hard core case, 
Dyson estimates the {\it minimum} (over all $\X$) of
${\cal W}_U(\X)$, for a $U$ similar to (\ref{yooo}), and gets a
lower bound for all $\rho$,  but 14 times smaller than (\ref{rho}). 
We follow another route.
An  important quantity for us will be the {\it average }
value of ${\cal W}_U(\X)$ in a cell, 
denoted by $\langle ~{\cal W}_U ~\rangle$.

To compute $ \langle  ~ U (\x_1 - \x_{j(1)})
~ \rangle$, for example it is easiest to do the $\x_2 ,
\ldots , \x_n$ integrations over the cell first and then the $\x_1$
integration. Provided $\x_1$ is in the smaller cube which is a distance
$R$ from the cell boundary (whose volume is $(\ell-2R)^3$),
the probability that $R_0 < |\x_j-\x_1| < R$
is $4 \pi (R^3 -R_0^3)/3\ell^{3} $.  
Thus, performing the $\x_1$ integration
over the smaller cube, and then adding similar contributions from
$U(\x_2-\x_{j(2)})$, etc., and using (\ref{yooo}),  we get
\begin{eqnarray}
\langle ~{\cal W}_U ~ \rangle
& \geq & \frac{3n(\ell -2R)^3}{(R^{3}-R_0^3)\ell^3}\left[ 1-\left( 1-
Q  \right)^{n-1} \right]
\label{avone} \\
& \geq & \frac{4\pi}{\ell^3} n (n-1)
\left(1 -\frac{2R}{\ell} \right)^3 \frac{1}{1+ Q(n-1)} .
\label{avtwo}
\end{eqnarray}
with 
\begin{equation}
Q= 4 \pi (R^3 -R_0^3) /3\ell^{3}\ll 1.
\end{equation}
In (\ref{avtwo}) we
used  $[1-x]^{n-1} \leq [1+(n-1)x]^{-1}$ for
$0\leq x\leq 1$. Note that (\ref{avtwo}) is of the form (\ref{provis}).

By similar reasoning, we obtain the upper  bound
\begin{eqnarray}
\langle ~{\cal W}_U ~ \rangle 
& \leq & \frac{3n}{R^{3}-R_0^3}\left[ 1-\left( 1-
Q \right)^{n-1} \right]
\label{avoneupper} \\
& \leq & \frac{4\pi }{\ell^3}n (n-1).
\label{avtwoupper}
\end{eqnarray}

Since $U(r)^2 = 4\pi (Q\ell^3)^{-1} U(r)$, we also obtain 
\begin{equation}
\langle~{\cal W}_U^2 ~ \rangle  \le 4\pi n~ (Q\ell^3)^{-1}
\langle~{\cal W}_U ~ \rangle .
\label{square}
\end{equation}

We can now use Lemma 1 and these 
averages to obtain (\ref{provis}) and (\ref{kay}). Instead of using 
(\ref{lower}) alone, we
pick some $0<\varepsilon \ll 1$ and, borrowing a bit of kinetic energy,
define
\begin{equation}
\widehat H \equiv \varepsilon {\cal T} +
( 1 - \varepsilon)~ \mu a   {\cal W}_U (\X) ~ .
\label{hamtwo}
\end{equation}
By Lemma 1 and $v\ge 0$, we have 
\begin{equation}
H > \widetilde H > \widehat H
\end{equation}
We shall derive (\ref{provis}) and (\ref{kay}) from a lower 
bound to  $ \widehat H$.

Although $\varepsilon$ is small, we regard
$H_0\equiv \varepsilon {\cal T}$ as our unperturbed Hamiltonian and
$V\equiv ( 1 - \varepsilon) \mu a   {\cal W}_U (\X)$ as a 
perturbation of $H_0$.
The ground state wave function for $H_0$ is $\Psi_0(\X) 
= \ell^{-3n/2}$ and  $H_0\Psi_0 = \lambda_0 \Psi_0=0$ 
(Neumann conditions). The second eigenvalue of $H_0$
is $\lambda_1 =\varepsilon \mu 
\pi/ \ell^2$.
Note that the ground state expectation, $\langle \Psi_0 |
 {\cal W}_U |\Psi_0 \rangle$, is precisely the average  
$\langle~{\cal W}_U ~ \rangle$ mentioned in (\ref{avone})--(\ref
{square}).

Temple's inequality \cite{TE} states that when a perturbation $V$ is
non-negative (as here) and when $\lambda_1-\lambda_0 \ge \langle \Psi_0
|V|\Psi_0 \rangle$ then the g.s.e., $E_0$, of the perturbed Hamiltonian
$H=H_0+V$ satisfies 
\begin{equation}
\!E_0 \ge \lambda_0\! +\! \langle \Psi_0|V|\Psi_0 \rangle \!-
\frac{ \langle \Psi_0
|V^2|\Psi_0 \rangle \!-\!\langle \Psi_0
|V|\Psi_0 \rangle^2    }{\lambda_1-\lambda_0 -\langle \Psi_0
|V|\Psi_0 \rangle }.
\label{temple}
\end{equation}

We apply this to our case  with $\lambda_1-\lambda_0 = \varepsilon \mu
\pi/ \ell^2$ and $V= (1-\varepsilon) \mu a {\cal W}_U$.  We neglect the
(positive) term $\langle \Psi_0 |V|\Psi_0 \rangle^2$ in (\ref{temple})
and we use (\ref{avtwo}), (\ref{avtwoupper}) and  (\ref{square}). 
We also use $1-\varepsilon < 1$ in two appropriate places and find 
\begin{equation}
\! \frac{E_0(n,\ell)}{\mu a \langle~{\cal W}_U~\rangle   } \ge
(\!1-\! \varepsilon)\!  
\left( 1- \frac{4\pi  a n}
{Q~\ell} \frac{  1}{ \varepsilon \pi - a\ell^2\langle~{\cal W}_U~\rangle}
\right).
\label{tempbound}
\end{equation} 

Apart from some higher order errors, (\ref{tempbound}) is just what we
need in (\ref{provis}), (\ref{kay}). Let us denote the order of the
main error by $Y^{\alpha}$, and we would like to show that $\alpha =
1/17$ suffices.  The errors are the  following:

{}From the $(1-\varepsilon)$ factor, we need $\varepsilon 
\le O(Y^{\alpha})$.

{}From the $Q(n+1)$ error in (\ref{avtwo}) we need $Q\rho \ell^3 
\le O(Y^{\alpha})$.

{}From the $R/\ell$ error in (\ref{avtwo}) we need $R/\ell \le 
O(Y^{\alpha})$.

{}From (\ref{tempbound}) we need  
$a\ell^{2} \langle~{\cal W}_U~\rangle /\varepsilon \le O(Y^{\alpha})$
and 
$ \rho \ell^5 a/(R^3
\varepsilon) \le O(Y^{\alpha})$.

All these desiderata can be met with $\varepsilon=Y^{\alpha}$,
$R/\ell = Y^{\alpha}$, $Q=O(Y^{\alpha})$,
$\rho R^3 =Y^{2\alpha}$
and $\alpha = 1/17$ --- as claimed. 

The partial support of U.S. National Science Foundation grant
PHY95-13072A01 (EHL) and the Adalsteinn Kristjansson Foundation of the 
University of Iceland (JY) is gratefully acknowledged.


\end{document}